\documentclass[reprint, twocolumn,amsmath,amssymb,aps,prl, showpacs, superscriptaddress]{revtex4-1}
\usepackage[utf8]{inputenc}
\usepackage{graphicx}% Include figure files
\usepackage{dcolumn}% Align table columns on decimal point
\usepackage{bm}% bold math
\usepackage[usenames]{color}
\usepackage{natbib}

\begin{document}

\newcommand{\comment}[1]{\textcolor{red}{ #1}}
\newcommand{\com}[2]{\begin{minipage}{0.48\textwidth}{\it #1}$\rightarrow$\textcolor{blue} {#2}\end{minipage}}
\newcommand{\co}[1]{{\it\textcolor{blue}{ #1}}}
\newcommand{\mco}[1]{{\it\textcolor{magenta}{ #1}}}
\newcommand{\JC}[1]{{\it\textcolor{red}{ #1}}}

\title{An acoustic analog to the dynamical Casimir effect in a Bose-Einstein condensate}

\author{J.-C.~Jaskula}
\altaffiliation[Present address: ]{Harvard-Smithsonian Center for Astrophysics, Cambridge, Massachusetts 02138, USA}
\author{G.~B.~Partridge}
\altaffiliation[Present address: ]{Agilent Laboratories, Santa Clara, CA 95051, USA}
\author{M.~Bonneau}
\author{R.~Lopes}
\author{J.~Ruaudel}
\author{D.~Boiron}
\author{C.~I.~Westbrook}
\affiliation{Laboratoire Charles Fabry, Institut d'Optique, CNRS, Univ Paris-Sud, 2 avenue Augustin Fresnel, 91127 Palaiseau France}

\date{\today}

\begin{abstract}
We have modulated the density of a trapped Bose-Einstein condensate by changing the trap stiffness, thereby modulating the speed of sound.
We observe the creation of correlated excitations with equal and opposite momenta, and show that for a well defined modulation frequency, the frequency of the excitations is half that of the trap modulation frequency. 
\end{abstract}

\maketitle

Although we often picture the quantum vacuum as containing virtual quanta whose observable effects are only indirect, 
it is a remarkable prediction of quantum field theory that the vacuum can generate real particles when boundary conditions are suddenly changed \cite{Moore1970,Fulling1976,Dodonov:10, Nation2011}. 
Known as the dynamical Casimir effect, a cavity with accelerating boundaries generates photon pairs. 
Recent experiments have demonstrated this effect in the microwave regime using superconducting circuits \cite{Wilson:11,Lahteenmaki:11}. 
Hawking radiation \cite{Hawking1974} is another situation characterized by spontaneous pair creation and 
work on sonic analogs to the Hawking problem \cite{Unruh1981} has led to the realization that Bose-Einstein condensates (BEC) are attractive candidates to study such analog models \cite{Garay2000,Balbinot2008,Lahav:10}, because their low temperatures promise to reveal quantum effects.
Here we exhibit an acoustic analog to the dynamical Casimir effect by modulating the speed of sound in a BEC. We show that correlated pairs of elementary excitations, both phonon-like and particle-like, are produced, in a process that formally resembles parametric down conversion \cite{Carusotto:10,Nation2011}.

The first analyses of the dynamical Casimir effect considered moving mirrors, but it has been suggested that a changing index of refraction could mimic the effect \cite{Yablonovitch:89,Dezael:10}. 
Our experiment is motivated by a suggestion in Ref.~\cite{Carusotto:10} 
that one can realize an acoustic analog to the dynamical Casimir effect 
by changing the scattering length in an interacting Bose gas.
The change in the interaction strength is analogous to an optical index change: the speed of sound (or light) changes. 
Seen in a more microscopic way, the ground state of such a gas is the vacuum of Bogoliubov quasi-particles whose makeup is interaction-dependent.
Changing the interaction strength projects this old vacuum onto a new state containing
pairs of the new quasi-particles~\cite{Carusotto:10}, which appear as pairwise excitations.
Instead of changing the interaction strength, we have simply modified the confining potential, which in turn changes the density.
Sudden changes such as these have also been suggested as analogs to
cosmological phenomena \cite{Fedichev:04, Jain:07, Hung:12}.

We study two situations, in the first the confining potential is suddenly increased 
and in the second the potential is modulated sinusoidally. 
The sinusoidal modulation of the trapping potential was studied in Ref.~\cite{Engels:07,Staliunas:04,Kagan:07} in the context of the observation of Faraday waves. 
Our results on sinusoidal modulation are similar to this work and we have
extended it to observe correlated pairs of Bogoliubov excitations. 
We produce these excitations in both the phonon and particle regimes, 
and observe correlations in momentum space. 
%Unlike the proposal of Ref.~\cite{Carusotto:10}, we observe correlations in momentum space rather than in position space. 
Parametric excitation of a quantum gas was also studied in optical lattices in which the optical lattice depth was modulated \cite{Schori:04,Tozzo:05}, although in that experiment, the excitation was observed as a broadening of a momentum distribution.

%%%%%%%%%%%%
% Experiment
%%%%%%%%%%%%
\begin{figure*}[t]
\begin{center}
\vskip 12pt
\includegraphics[width=16cm]{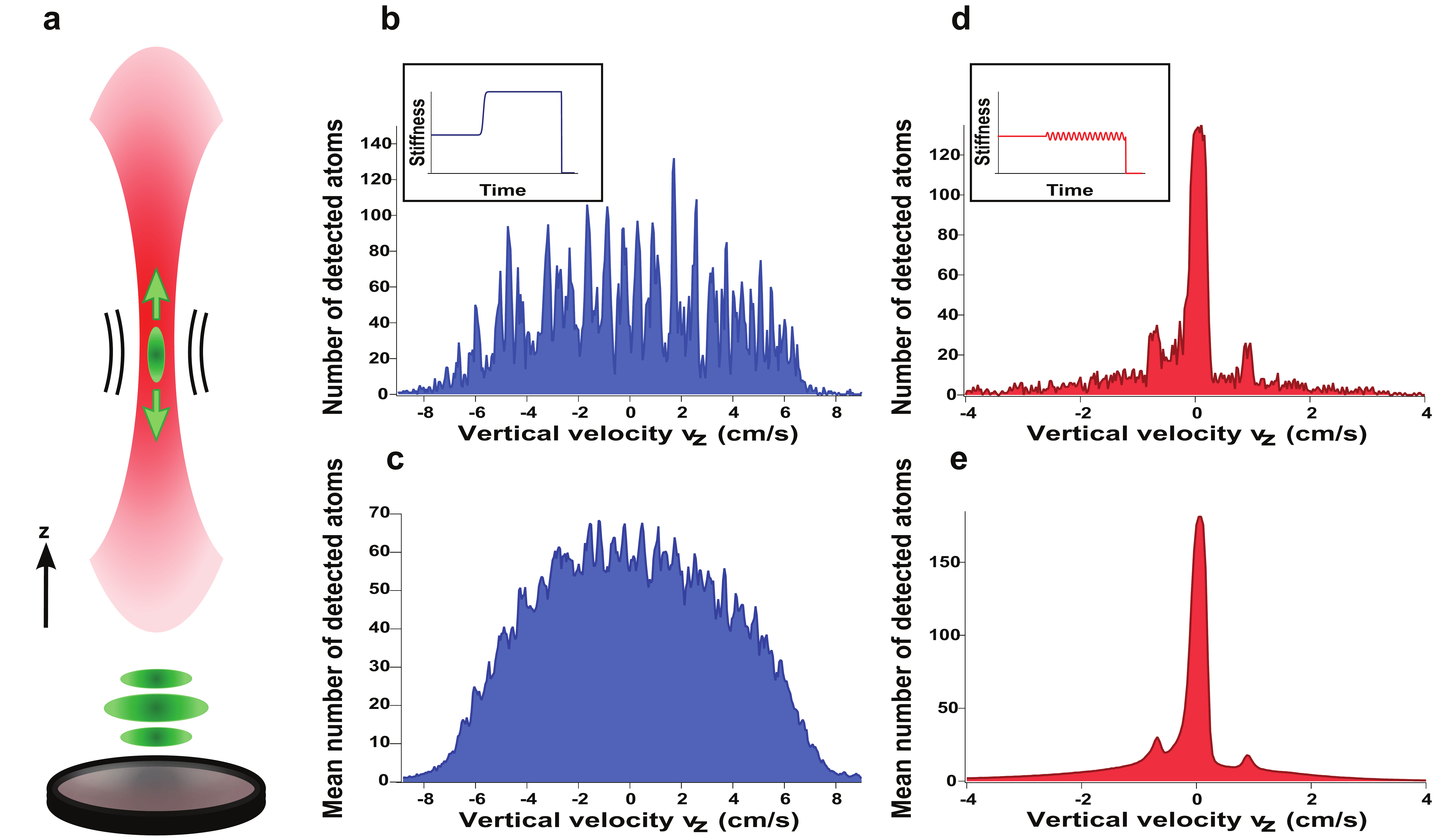}
\caption {Effects of time-varying potentials (color online). \textbf{a}, Schematic view of the experiment. Pairs of Bogoliubov quasiparticles are created by varying the trap stiffness. 
After the flight to the detector these excitations appear as a broadening or sidebands on the atom cloud in the vertical ($z$) direction. 
In the following plots we convert arrival times to relative velocities 
and average over the transverse dimensions. 
\textbf{b,}
Single shot velocity distribution for a cloud which was subjected to a sudden increase in the trap stiffness. 
The inset shows the time evolution of the trap stiffness. \textbf{c,}
As in \textbf{b} but averaged over 50 shots.
\textbf{d,}
Single shot velocity distribution for a cloud which was subjected to a weak, sinusoidal modulation of the trap stiffness at 2.17~kHz. 
The inset shows the time evolution of the trap stiffness.
\textbf{e,} As in \textbf{d} but averaged over 780 shots. 
}
\label{fig:introGen}
\end{center}
\end{figure*}

The experimental apparatus is the same as that described in Refs.~\cite{Partridge:10,Jaskula:10} and is shown schematically in Fig.~\ref{fig:introGen}a.
We start from a BEC of approximately $10^5$ metastable helium (He*) atoms evaporatively cooled in a vertical optical trap to a temperature of about 200 nK.
The trapped cloud is cigar shaped with axial and radial frequencies
of 7~Hz and 1500~Hz. 
In the first experiment we raise the trapping laser intensity by a factor of 2 with a time constant of 50~$\mu$s using an acousto-optic modulator (see inset to Fig.~\ref{fig:introGen}b).
The trap frequencies thus increase by $\sqrt 2$.
The compressed BEC is held for 30 ms before the trap laser is switched off (in less than 10 $\mu$s). The cloud falls onto a position sensitive, single atom detector which allows us to measure the
atom velocities\cite{Supplementary}.
After compression, the gas is excited principally in the vertical direction: transversely we only observe a slight heating (about 100 nK). 
Figure~\ref{fig:introGen}b shows a single shot distribution of vertical atom velocities relative to the center of mass and integrated horizontally, 
while Fig.~\ref{fig:introGen}c shows the same distribution averaged over 50 shots. 
These distributions are more than one order of magnitude wider than that of an unaffected BEC.
The individual shots show a complex structure which is not reproduced from shot to shot, as is seen from the washing out of the peaks upon averaging.

\begin{figure}[t]
\begin{center}
\includegraphics[width=\columnwidth]{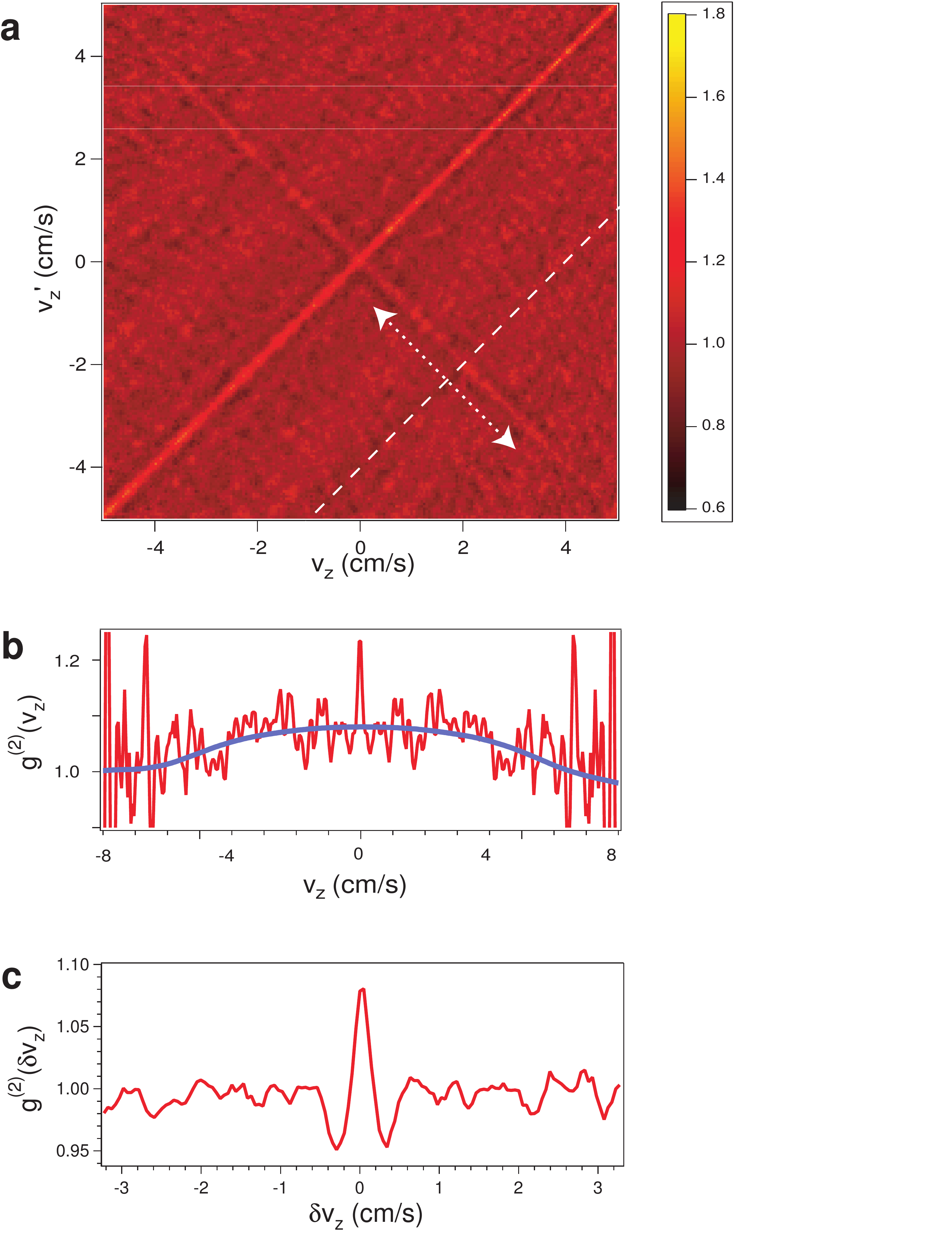}
\caption {Density correlations after a sudden compression (color online). \textbf{a,} Normalized correlation function $g^{(2)}(v_z,v_z')$ of the data in Fig.~\ref{fig:introGen}\textbf{c} (50 shot average).  
The signal on the diagonal results from the density fluctuations in the cloud. 
The anti-diagonal line indicates the creation of correlated quasiparticles with opposite momenta,
and is the signature of the dynamical Casimir effect. 
\textbf{b,} Anti-diagonal correlation function $g^{(2)}(v_z,v_z^\prime=-v_z)$. 
The smooth line shows the result of smoothing the data over a window of about 1 cm/s.
The correlations apparently persist over a scale comparable to that of the density distribution. \textbf{c,} Correlation function along the dashed line and integrated over a region indicated by the dotted arrows, as a function of $\delta v_z = v^\prime_z -v_z$.
The dips on either side the peak may be related to the effect reported in \cite{Imambekov:2009}.
}
\label{fig:2DSudden}
\end{center}
\end{figure}

We consider the correlations between atoms with vertical velocities $v_z$ and $v_z^\prime$, 
by constructing a normalized second-order correlation function, $g^{(2)}(v_z,v_z^\prime$)\cite{Supplementary}, averaged over the $x$-$y$ plane and shown in Fig.~\ref{fig:2DSudden}a. 
The plot exhibits two noticeable features along the $v_z^\prime=v_z$ and $v_z^\prime=-v_z$ diagonals.
The former reflects the fluctuations in the momentum distribution, as in the Hanbury Brown and Twiss effect \cite{Schellekens:05}, except that this cloud is far from thermal equilibrium.
The $v_z^\prime=-v_z$ correlation is a clear signature of a correlation between quasi-particles of opposite velocities. 
A projection of this off-diagonal correlation is shown in Fig.~\ref{fig:2DSudden}b.
At low momentum, the excitations created by the perturbation are density waves (phonons)
which in general consist of superpositions of several atoms traveling in opposite directions.
In the conditions of our clouds, a phonon is adiabatically converted into a single atom of the same momentum during the release by a process referred to as ``phonon evaporation" \cite{Tozzo:04}.
Therefore in the phonon regime as well as in the particle regime, 
we interpret the back-to-back correlation in Fig.~\ref{fig:2DSudden}a as the production of pairs of Bogoliubov excitations with oppositely directed momenta as predicted in the acoustic dynamical Casimir effect analysis \cite{Carusotto:10}.

\begin{figure}[tb]
\begin{center}
\vskip 12 pt
\includegraphics[width=6.5 cm]{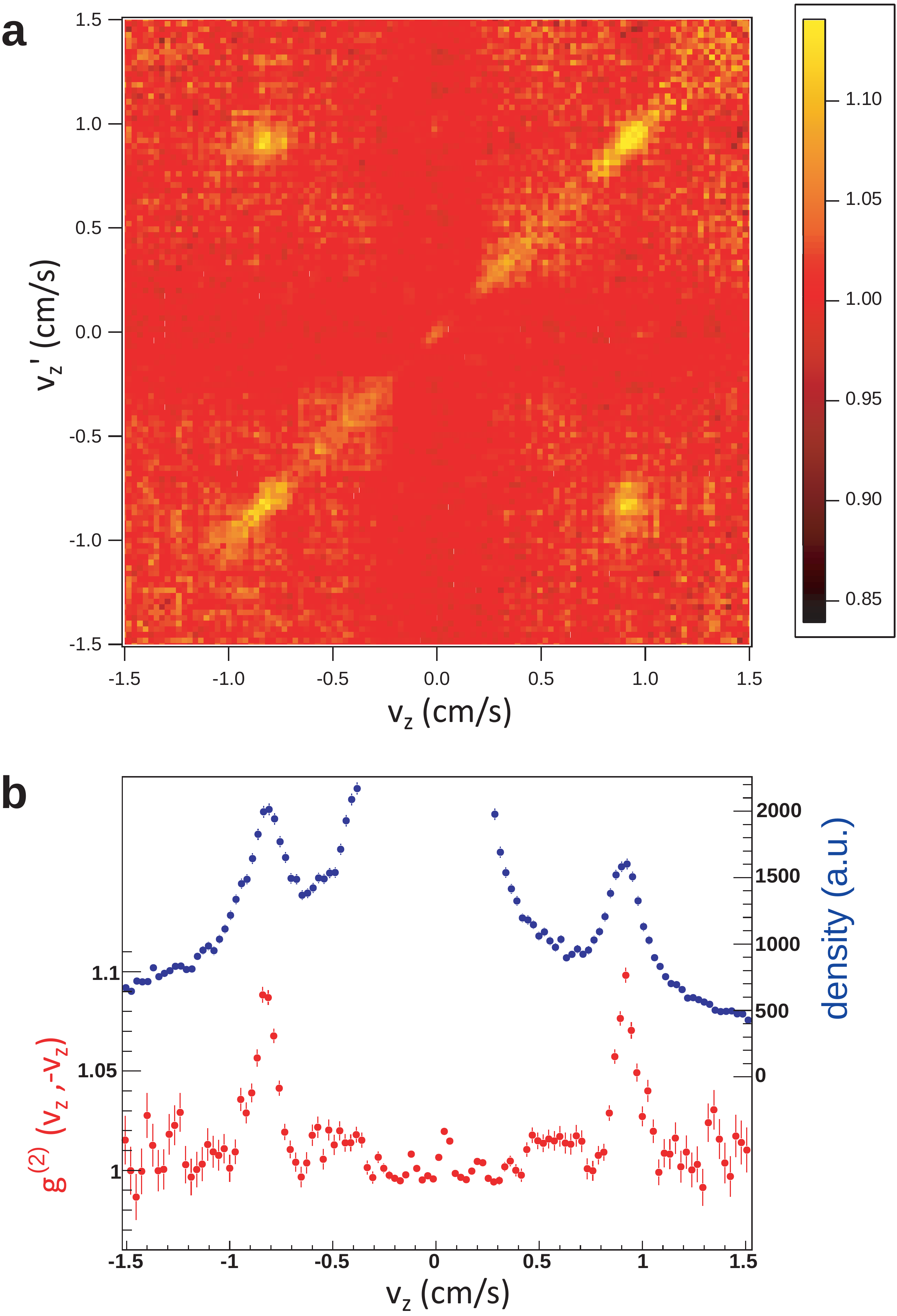}
\caption{Density correlations after a periodic modulation (color online). \textbf{a,} Normalized correlation function $g^{(2)}(v_{z},v'_{z})$ measured after sinusoidal modulation of the trap frequency at a frequency $\omega_\mathrm{\mathrm m}/2 \pi=2.17\,\rm{kHz}$, averaging over 243 experimental shots. We observe a strong correlation between well defined, oppositely directed velocities. \textbf{b,} 
Plot of the density distribution (blue) and of the anti-diagonal velocity correlation function, 
 $g^{(2)}(v_z,v_z^\prime=-v_z)$ (red). }
\label{fig:2DMonoFreq}
\end{center}
\end{figure}

To further study this process, we replace the compression by a sinusoidal modulation of the laser intensity $I(t)=I_0(1 + \delta  \cos \omega_{\mathrm m} t)$ 
(inset of Fig. \ref{fig:introGen}d).
We choose $\delta$ such that the trap frequencies are modulated peak to peak by about 10\%.
The modulation is applied for 25~ms before releasing the condensate.
Figures~\ref{fig:introGen}d and \ref{fig:introGen}e show
respectively single shot and averaged momentum distributions resulting from the modulation.
One sees that the momentum distribution develops sidebands, approximately symmetrically placed about the center. 
Figure~\ref{fig:2DMonoFreq}a shows the normalized correlation function, 
plotted in the same way as in Fig.~\ref{fig:2DSudden}a, for a modulation frequency $ \omega_{\mathrm m} /2 \pi = 2170$~Hz. 
We again observe anti-diagonal correlations as for a sudden excitation except that the correlations now appear at a well defined velocity, which coincides with that of the sidebands (see Fig.~\ref{fig:2DMonoFreq}b).

\begin{figure}[t]
\begin{center}
\includegraphics[width=\columnwidth]{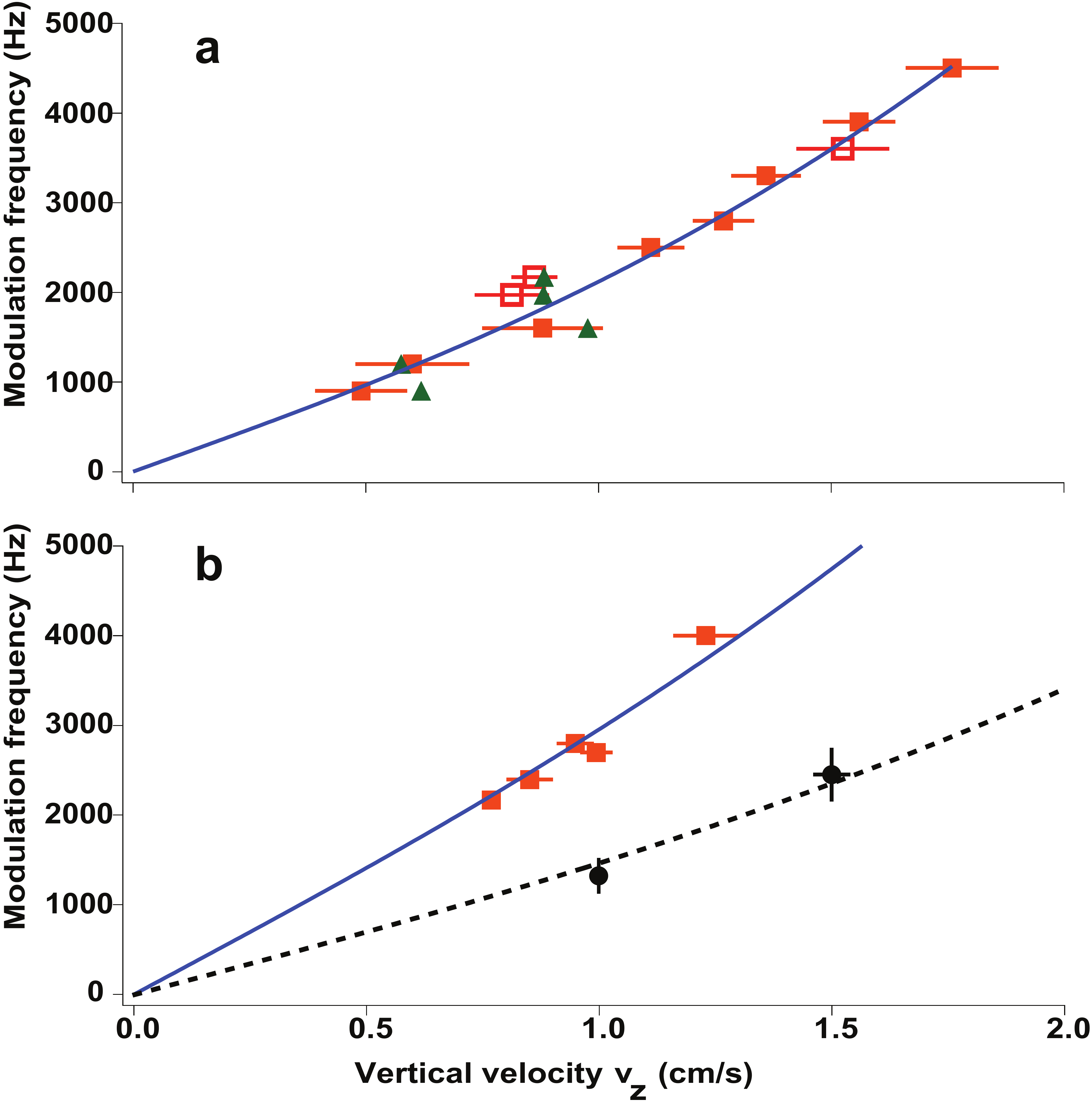}
\caption
{Dispersion relation observed by modulating the trap depth (color online). 
\textbf{a,} The orange squares show the sideband velocity determined from the density distributions.
The green triangles are derived from the correlation functions of the same data. 
The curve is a fit to the dispersion relation \eqref{math:BogPDC} as described in the text.
Only the solid squares were included in the fit: these points were all taken on the same day, whereas the open squares were taken under slightly different trap conditions, with possibly different density. 
The error bars are statistical estimates based on the fits to the velocity distributions such as in figure~\ref{fig:2DMonoFreq}b.
\textbf{b,} Comparison between trap modulation and Bragg scattering. 
The black circles are observations of the dispersion relation by Bragg spectroscopy.
The orange squares are found as in (a), and clearly show that the corresponding frequency is 
about a factor of two higher than in the Bragg data at a given velocity. 
The curves show the two fits discussed in the text.
The vertical error bars on the Bragg data are determined by fits to the Bragg resonances.
} 
\label{fig:dispersion}
\end{center}
\end{figure}

We have examined sinusoidal modulation for frequencies
$\omega_{\mathrm m}/2 \pi$ between $900\,\rm{Hz}$ and $5000\,\rm{Hz}$ and observed excitations similar to those in Fig.~\ref{fig:2DMonoFreq}. 
We summarize our observations in Fig.~\ref{fig:dispersion}a in which we plot the excitation frequency as a function of the sideband velocity.
We also plot the locations of the peaks in the correlation functions on the same graph.  
For modulation frequencies much above 2 kHz, the antidiagonal correlation functions are quite noisy preventing us from clearly identifying correlation peaks. 
This noise may have to do with the proximity of the parametric resonance with the
transverse trap frequency ($\sim 3$~kHz)~\cite{Staliunas:04}.
 
A weakly interacting quantum gas obeys the well known Bogoliubov-de Gennes dispersion relation between the frequency $\omega_{\mathbf k}$ and wavevector $k$: 
\begin{equation}
\label{math:BogPDC}
\omega_\mathbf{k}= \alpha \sqrt{c^2 k^2+\left({\hbar k^2\over 2m}\right)^2}
\end{equation}
with $\alpha = 1$ and $c$, the sound velocity.
This relation describes both phonons (long wavelength excitations) whose dispersion
is linear and free particles, whose dispersion is quadratic.
If our observation indeed corresponds to the creation of pairs,
we expect the total excitation energy to be shared between the two excitations.
Momentum conservation, on the other hand, requires that the two energies be equal, implying
$\omega_\mathrm{m} =2 \omega_{\bf k}$. 
Therefore the relation between the modulation frequency and the sideband
velocity should also be given by Eq.~\ref{math:BogPDC} but with $\alpha=2$ and $k = m v_z/\hbar$. 
Fitting the points in Fig.~\ref{fig:dispersion}a to \eqref{math:BogPDC} with $\alpha$ and $c$ as free parameters, we obtain $\alpha=2.2 \pm 0.3$.
The fitted sound velocity, $8\pm 3$~mm/s, is consistent with the value one can calculate from the trap parameters and the estimated number of atoms \cite{Supplementary}.

We can further corroborate our interpretation of pairwise excitations by a method more direct and robust than the 2 parameter fit to the data in Fig.~\ref{fig:dispersion}a.
In Fig.~\ref{fig:dispersion}b, we compare the dispersion relation resulting from modulation with that obtained by Bragg scattering.
Bragg scattering produces single excitations of quasiparticles at a definite energy and momentum \cite{Ozeri:05}.  
We excited the BEC with two lasers in the Bragg configuration to determine the frequency for a given $k$-vector \cite{Supplementary}.
Then, under the same experimental conditions, using sinusoidal trap laser modulation, we excited the BEC at various frequencies and found the corresponding velocities.
The lower curve in Fig.~\ref{fig:dispersion}b is a fit to the Bragg data in which we fix $\alpha=1$ and fit the speed of sound.
The upper curve is a fit to the trap modulation data in which we set the speed of sound to that found in the first fit and we allowed $\alpha$ to vary. 
This second fit yields $\alpha=2.07\pm 0.2$. 
The fitted speed of sound for this data set (about 13 mm/s) is higher than in the data of Fig.~\ref{fig:dispersion}a, 
because during these runs the number of atoms in the condensate was larger.

An even more dramatic confirmation of our interpretation would be the observation of
sub-Poissonian intensity differences in the two sidebands, 
as was observed in the experiment of Ref.~\cite{Wilson:11}, as well as in
Refs.~\cite{Buecker2011}.
The latter experiment modulated the center of a trapped, one dimensional gas producing transverse excitations which in turn produced twin beams.
Equivalently, one could ask whether the Cauchy-Schwarz inequality is violated \cite{Kheruntsyan:12}, indicating a non-classical correlation.
Comparing intensity differences in the sidebands we observe a reduction of the fluctuations compared to uncorrelated regions of the distribution. 
However, we observe no sub-Poissonian fluctuations or Cauchy-Schwarz violation,
probably because of a background under the sidebands (see Fig.~\ref{fig:introGen}d). 
The background is due to atoms spilling out of the trap before release.

Another difference between our experiment and an ideal realization of the dynamical
Casimir effect is that the temperature is not negligible. 
This means that the pair generation did not arise from the vacuum but
rather from thermal noise. 
For our temperature of 200~nK, 
the thermal occupation of the mode of frequency 2~kHz is 1.6.
In the absence of a thermal background, the normalized correlation function would show an even higher peak.
Using the perturbative approach of Ref.~\cite{Carusotto:10}, one can show that  $g^{(2)}(v_z,v_z^\prime=-v_z$) is a decreasing function of the temperature, since thermal quasi-particles are uncorrelated and only dilute the correlation.

Many authors have discussed the relationship of the dynamical Casimir effect to Hawking and Unruh radiation (see \cite{Nation2011} for a recent review).
It has also been pointed out that the two-particle correlations
arising in the sonic Hawking problem constitute an important potential detection strategy
\cite{Balbinot2008,Larre:12},
although the above authors discussed correlations in position space.
The present study has confirmed the power of correlation techniques, and shown in addition that \emph{momentum space} is a good place to look for them.
We expect that a similar approach can be applied to
Hawking radiation analogs as well as the general problem of studying the physics of curved spacetime by laboratory analogies.

\vskip 6pt
\begin{acknowledgments}
This work was supported by the IFRAF institute, the Triangle de la Physique, the ANR-ProQuP project, J.R. by the DGA, R.L. by the FCT scholarship SFRH/BD/74352/2010, and GBP by the Marie Curie program of the European Union.
We acknowledge fruitful discussions with D. Cl\'ement, I. Carusotto, A. Recati, R. Balbinot, A. Fabbri, N. Pavloff and P.-E. Larr\'e.
\end{acknowledgments}

\vskip 12pt
\noindent {\large{\textbf{Supplemental material}}}
\vskip 6pt

\noindent \textbf{Detection and data analysis.}
The atoms are detected after falling 46.5 cm to a position sensitive detector which allows reconstruction of the arrival time and horizontal position of individual atoms \cite{Schellekens:05}. The mean time of flight is 307 ms. Given the value of the vertical Thomas-Fermi radius (0.5 mm), this time is long enough for the arrival time to reflect the vertical velocity, provided this velocity is well above 1.5 mm/s.

The experimental correlation function corresponds to a 2D-histogram of the vertical velocities $v_z$ and $v_z^\prime$ of each possible pair of atoms originating from a single condensate, and averaging over all realizations. 
For normalization we divide this first histogram by a second 2D-histogram obtained with pairs of atoms originating from separate realizations. 
We observe shot to shot fluctuations in the arrival times and widths of the condensate. 
To avoid spurious correlation signals arising from fluctuations in arrival time, we recentered each shot by aligning the condensate peaks before averaging and normalizing.
The fluctuations in the width were reduced by selecting two different width classes and normalizing them separately before adding them to obtain Fig.~3.

\noindent \textbf{Speed of sound in a quasi-condensate.}
For our typical atom number and trap frequencies, the atomic clouds are in the 1D-3D cross-over. We follow the model of Ref.~\cite{Gerbier:04} to calculate the speed of sound of atoms confined in a cylindrical trap and obtain $mc^2=\frac{\hbar\omega_\bot}{2}\left(\tilde\mu-1/\tilde\mu\right)$ where $\omega_\bot/2\pi$ is the transverse frequency of the trap and $\tilde\mu>1$ the chemical potential in units of $\hbar\omega_\bot$. We then apply a local density approximation for the longitudinal confinement on the dynamical structure factor \cite{Ozeri:05} to relate the speed of sound of equation (1) with the atom number.

\noindent \textbf{Bragg spectroscopy.} In the Bragg spectroscopy measurement of Fig.~4,
%\ref{fig:dispersion}b, 
we use two laser beams at an angle $\theta$ to provide an excitation wavevector $\mathbf{k}=2\times \sin{(\theta/2)} k_{\mathrm laser}\mathbf{e_z}$. 
The angles were approximately $6^\circ$ and $9^\circ$. 
The wave vector was measured by applying an intense 10~$\mu$s pulse which populated many diffraction orders. 
The positions of the diffraction orders permitted an accurate fit to find the wavevector.
For Bragg diffraction, the excitation pulse was of 5 ms duration. 
We then varied the relative frequency of the two laser beams to find the Bragg resonances for both positive and negative frequency differences. 
The difference in the position of the resonances divided by 2 was used as the frequency 
$\omega_{\mathbf k}$.
The Bragg spectra exhibit a single sideband, showing that phonon excitations appear with a single momentum, as predicted by the phonon evaporation scenario.

\bibliographystyle{apsrev}
%\bibliography{DC_prl}

\end{document}